# Resonant Spin-Transfer-Driven Switching of Magnetic Devices Assisted by Microwave Current Pulses

Y.-T. Cui, J. C. Sankey, C. Wang, K. V. Thadani, Z.-P. Li, R. A. Buhrman, D. C. Ralph*

*Cornell University, Ithaca, New York 14853, USA*

The torque generated by the transfer of spin angular momentum from a spin-polarized current to a nanoscale ferromagnet can switch the orientation of the nanomagnet much more efficiently than a current-generated magnetic field, and is therefore in development for use in next-generation magnetic random access memory (MRAM).  Up to now, only DC currents and square-wave current pulses have been investigated in spin-torque switching experiments.  Here we present measurements showing that spin transfer from a microwave-frequency pulse can produce a resonant excitation of a nanomagnet and lead to improved switching characteristics in combination with a square current pulse.  With the assistance of a microwave-frequency pulse, the switching time is reduced and achieves a narrower distribution than when driven by a square current pulse alone, and this can permit significant reductions in the integrated power required for switching.  Resonantly excited switching may also enable alternative, more compact MRAM circuit architectures.

A spin-polarized current can exert a spin-transfer torque[1,2] on a nanomagnet strong enough to reverse its magnetization without an applied magnetic field.[3,4] This provides a promising mechanism for writing information in the next generation of magnetic random access memory (MRAM). Previous studies have indicated that the spin-torque switching proceeds via a process of magnetic precession with increasing precession amplitude.[5-8] This suggests that using a high-frequency (RF) current to drive precession on resonance[9,10] might improve the switching characteristics.[11] Here we report low temperature proof-of-principle experiments, using a combination of RF and square-wave current pulses, which show that resonantly-excited precession can enhance the switching speed and the reproducibility of switching times for spin-torque-driven nanoscale magnetic devices, and also significantly reduce the integrated power required for switching.

The initial experiments on spin transfer switching studied primarily multilayered nanopillar devices (Fig. 1a inset) in which the magnetic moment directions were approximately collinear for the "free" magnetic layer that undergoes switching and the "pinned" magnetic layer that polarizes the current.[12-14] Since the spin torque vanishes at zero offset angle, spin-transfer switching in such devices requires that the moments have an initial deviation from completely parallel or anti-parallel alignment that was usually initiated by thermal fluctuations. The switching process was therefore stochastic, with a broadened distribution of switching times. An alternative strategy is to start with a non-zero equilibrium angle between the two moments by biasing one of the two layers away from the easy axis defined by shape anisotropy.[6,15-17] However, in this case using a square-wave current pulse (with a single sign of current) to drive switching may not provide optimum efficiency. Consider the initial stage of the switching process in a thin film nanopillar device, where the precession amplitude of the free-layer moment is less than the offset angle between the two moments (Fig. 1b). For positive currents (by our convention) the spin torque acts to push the free layer moment away from the pinned moment. This means that in one half of the precession cycle the spin torque from a square current pulse works to increase the precession amplitude, but in the other half of the cycle it acts to decrease the amplitude. In contrast, the alternating signs of torque from an RF current at the precession frequency can act always in the direction to increase the precession amplitude throughout the entire precession cycle. Therefore, RF excitation can be expected to be more efficient than square current pulses for small precession angles in this geometry. After the precession amplitude has grown to exceed the offset angle, a square pulse will be preferred since its torque is then in the right direction to enhance precession throughout the whole precession cycle. In our work, we study the switching process driven by a combination of RF and square current pulses. We use an initial RF pulse to excite the free layer magnetization resonantly, and then we complete the switching with a square current pulse.

We study exchange-biased spin valve nanopillars with the layer structure (in nm) 8 IrMn / 4 permalloy / 8 Cu / 4 permalloy / 2 Cu / 30 Pt, with Cu contacts on top and bottom. Here permalloy (Py) is $Ni_{81}Fe_{19}$. All the layers are etched to have an elliptical cross section of approximately $80 \times 150$ nm$^2$. The long axis of the ellipse is designed to be 45° from the direction of the exchange bias between the IrMn layer and the bottom Py layer (pinned layer). Based on a macrospin-model fit to the curve of magneto-resistance as a function of the magnetic field applied along the exchange bias direction (Fig. 1a), we estimate that the actual equilibrium offset angle between the free and pinned layers in the low-resistance (LR) state is $\approx 30°$. We perform the majority of our measurements at a background temperature of 20 K to reduce thermal fluctuations. We estimate[18] that the samples may heat to ~ 100 K for our largest current pulses. During the switching measurements, a field $H_d = 140$ Oe is applied along the exchange bias direction to cancel the dipole field from the pinned layer on the free layer, leaving the free layer near zero net magnetic field. We will focus on switching from the LR state to high-resistance (HR) state. (Measurements of HR to LR switching are qualitatively similar.). Extrapolated to zero temperature, the switching current and switching field for LR to HR switching were $I_{c0} = 3.2 \pm 0.2$ mA and $|H_{c0} - H_d| = 260$ Oe.

Fig. 1c shows the type of pulse waveform we employ, consisting of a combination of RF and square-wave pulse signals chosen so that the switching process can be initiated by a RF pulse and then completed by the square pulse. Our methods for generating and calibrating such waveforms are described in the Supplementary Material. Positive current will correspond to electrons flowing from the free layer to the pinned layer. We split the excitation signal using a power divider, with one part directed to a 12.5 GHz oscilloscope for recording the waveform, and the other directed to the sample through the RF port of a bias-tee. One particularly-important parameter of the waveform is the phase of the RF signal at the onset time of the square pulse, which we call $\phi$. We define $\phi$ to be zero when the RF current would be crossing zero in the positive direction at the moment of the initial rise of the square pulse. Our uncertainty in $\phi$ is determined by a trigger jitter of 5 ps, and corresponds to $\Delta\phi = 9°$ for $f = 4.9$ GHz.

We begin by considering the regime in which the square pulse is sufficiently large (7.7 mA) and short (0.7 ns) so that thermal effects do not affect the switching trajectory dramatically (smaller, longer square pulses are discussed below). The amplitude of the square pulse is chosen to be just below the critical value at which the switching probability driven by the 0.7 ns square pulse alone becomes nonzero for this sample. Using a fixed RF peak amplitude (1.4 mA) and length (1.7 ns), we vary the RF frequency $f$ in discrete steps and at each value we vary the relative phase $\phi$ by more than 360 degrees. Fig. 1d shows the maximum switching probability that can be achieved at each value of $f$ after optimizing $\phi$. (For each point we average over 150 identical pulses.) We observe a strong resonant response to the RF pulse, in that the switching probability is enhanced from nearly 0 to over 90% at the resonant frequency $f_0 = 4.9$ GHz. The switching probability oscillates as a function of $\phi$ with a period of 360 degrees (Fig. 1e). At 4.9 GHz, the switching probability is greatest (92%) at $\phi = 50°$ and least (6%) at $\phi = 230°$. As $f$ is tuned through the resonance, the optimum value of $\phi$ varies by approximately 180°. In contrast to the strong dependence of the switching probability on the RF phase at the onset of the square pulse ($\phi$), there is negligible dependence on the initial phase at the start of the RF pulse.

Fig. 1f shows the effect of the RF pulse on the experimental distribution of switching probabilities as a function of square pulse duration, for $f = 4.9$ GHz, $\phi = 50°$, and RF pulse parameters of 1.4 mA peak amplitude and 0.3 ns duration prior to the square pulse. Each point corresponds to an average over 500 pulses. Compared to the distribution for no RF component, the length of the square pulse necessary to achieve switching is reduced, and the width of the distribution of switching times is narrowed dramatically, by a factor of 2.6. The length of the square pulse necessary to achieve switching with 95% probability changes from 1.64 ns to 0.93 ns, significantly more than the 0.3 ns duration of the RF pulse. We attribute the narrowing of the distribution to the coherent excitation of precession by the RF pulse, which may mitigate the effects of thermal fluctuations.

In order to visualize how an RF pulse affects the free layer precession, we have performed macrospin simulations including thermal fluctuations corresponding to a temperature of 20 K (Fig. 2). An example of the simulated dependence of switching probability on $\phi$ is shown in Fig. 2a. The simulations suggest that the free-layer magnetization becomes phase-locked[19] to the RF drive and precesses in-phase with it with increasing amplitude. At the end of the RF pulse, the square pulse is applied, producing a torque on the free layer pointing away from the pinned layer moment. For $\phi = 50°$ (Fig. 2b top), the square pulse begins just as the free layer moment enters the half of its precession cycle in which it is farther from the pinned-layer moment. By the time that the free layer precesses to the extremum position, the current from the square pulse increases to a large value and applies a strong torque to increase the precession amplitude of the free layer. For a sufficiently large precession angle, the spin torque can then act to enhance precession in the whole precession cycle and it can complete the switching in a few cycles. On the other hand for $\phi = 230°$ (Fig. 2b bottom), the free layer is in the half of its precession cycle closer to the pinned-layer moment at the time when square pulse is applied and initially the spin torque acts to decrease the precession amplitude. The result is a decreased switching probability compared to $\phi = 50°$.

Next we explore how varying the RF pulse parameters affects the switching probability. We first measure the frequency dependence of switching probability with the same parameters as in Fig. 1c except that we use several different values for the amplitude of the RF pulse (Fig. 3a). As the RF peak amplitude is increased from 0.7 mA to 1.8 mA, the maximum switching probability at the resonant frequency increases and eventually saturates. In Fig. 3b we show how the switching probability for $f = 4.9$ GHz depends on the duration of the RF pulse, for each of the RF amplitudes shown in Fig. 3a. In each case, the onset of the square pulse is timed to correspond to the end of the RF pulse, and the relative phase $\phi$ is optimized to maximize the switching probability. For all of the RF amplitudes, we observe that the probability saturates at long RF pulse durations, indicating that the precession amplitude excited by a given RF amplitude also saturates. Larger RF amplitudes give higher switching probabilities at saturation and reach saturation for shorter RF pulse durations. For the RF amplitude of 1.8 mA, the switching probability saturates for RF pulse durations as short as 300 ps, corresponding to only 1.5 cycles of RF current.

These results are in good accord with our macrospin simulations, which indicate that the RF drive causes the precession amplitude to grow with time until the point that the energy gained from spin transfer over one cycle is balanced by the energy lost to damping. Therefore, for small RF amplitudes, the saturation angle for the precession increases with RF amplitude. However, even for large RF amplitudes the precession amplitude cannot grow much beyond the offset angle between the free and pinned magnetic layers, because the spin-transfer torque changes sign to damp precession on the part of the magnetic trajectory where the precession angle exceeds the offset angle. It is possible that nonlinearities in the magnetic dynamics may also contribute to the saturation as a function of RF amplitude.[20]

In Fig. 3c, for a fixed square-wave amplitude of 7.7 mA, we compare the square pulse durations required to achieve 95% switching probability, $T_{0.95}$, as a function of RF amplitude at the resonance frequency ($f$ = 4.9 GHz) for both the optimum relative RF phase ($\phi$ = 50°) and the least-favorable phase for switching ($\phi$ = 230°). In all cases, the RF duration (1.7 ns) is long enough to achieve saturation. For the optimized value of $\phi$, $T_{0.95}$ for all RF amplitudes is considerably less than for switching driven by a square pulse alone (dashed line). For the least-favorable relative phase $\phi$ for switching, $T_{0.95}$ is close to the square-pulse-alone value over most of the range of RF amplitude, and for very small RF amplitudes the RF pulse can actually suppress the switching slightly. The potential for power savings that might be achieved by using resonant RF pulses is illustrated in Fig. 3d. This panel shows two excitation pulses, one with an RF component and one without, which both give identical 95% switching probabilities. The integrated power consumption needed for switching is reduced by 40% when RF pulse is used.

While our results provide a proof of principle that resonantly-enhanced switching can provide better performance than switching by square pulses alone, for most applications switching must be achieved reliably in the presence of thermal fluctuations at room temperature, in contrast to our low-temperature results. In order to understand the influence of thermal fluctuations, and whether enhanced switching with resonant RF pulses might be extended to room temperature, we have performed experiments using square pulses with a lower amplitude (4.8 mA) and longer duration (2 ns), for which thermal fluctuations are significant even for a 20 K background temperature, and we also performed room temperature studies. At 20 K, we employed 1.4 mA, 0.8 ns RF pulses, which are sufficient to saturate the RF-excited precession even though they have considerably less amplitude than the square pulse. We find that the switching probability is still enhanced when assisted by RF pulse but its variation as a function of the relative phase $\phi$ is much weaker (Fig. 4a) than for the case of a larger, shorter square pulse (Fig. 1e). In Fig. 4b we compare the distributions of switching times for a pulse with no RF component and for a pulse with an RF component with an optimized value of the relative phase $\phi$. With the RF pulse, the square pulse duration needed for switching is shorter, but the width of the distribution is comparable to the case with no RF pulse. These results suggest a crossover to a regime in which thermal fluctuations occurring during the square pulse begin to dominate the dynamics. The RF pulse can still enhance switching by giving the free layer an increased magnetic energy before the start of the square pulse, but the lack of strong phase dependence and the wide distribution of switching times indicate that the final switching process is dominated by thermal processes during the square pulse.

For temperatures above 100 K for our device geometry, we find that large square pulse amplitudes are necessary in order to observe any phase-dependent enhancement from an RF pulse -- large enough square-pulse amplitudes to drive switching by themselves in just a few precession cycles. Fig. 4c and 4d show the results of room temperature measurements using an RF pulse (1.4 ns, 1.5 mA) chosen to saturate the free layer precession, together with a 0.5 ns 10.8 mA square pulse. The distribution of switching times is centered at a slightly shorter square pulse duration when RF pulse is used, but the widths of the distributions in switching times are similar.

These findings suggest a strategy for how the device design might be optimized in an effort to extend the benefits of resonantly-enhanced switching to room temperature. For the value of the equilibrium offset angle between the free and pinned layer magnetizations in our samples ($\approx 30°$), we have seen that relatively small RF amplitudes are capable of exciting the precession amplitude to saturation (to a value close to the offset angle, according to macrospin simulations) irregardless of thermal effects. The thermal fluctuations appear to interfere with the switching process primarily while the much larger square pulses are applied to complete the switching. This suggests that improving reliability in the presence of thermal fluctuations might be achieved by using a device design which allows the more-effective RF pulse to control the magnetic dynamics over a larger angular range, so as to rely less on the square-pulse component of the waveform. We therefore propose using devices in which the pinned layer moment is exchange-biased at a very large angle, close to 90°, relative to the free-layer moment. It is true that this geometry will reduce the magnetoresistance signal used to read out the orientation of the free layer in memory devices, but the extraordinarily-large magnetoresistance values provided by MgO-based magnetic tunnel junctions[21] may provide sufficient read signal even for large offset angles, or else the free-layer orientation might be read out using a second junction on the opposite side from the pinned layer. An offset angle near 90° should allow phase-locking with the RF pulse to excite the free-layer switching coherently almost to the switching threshold, at which point only a modest square pulse may be needed to complete the switching process even in the presence of thermal fluctuations. This geometry also provides the advantage that an offset angle near 90° maximizes the strength of the RF torque per unit current, which may allow the use of thicker, more thermally-stable magnetic free layers without requiring larger RF current amplitudes. In addition to enabling potential improvements in switching speeds, reproducibility, and integrated power consumption, a successful implementation of resonant spin transfer switching might open new possibilities for storage architectures. For example, if the resonant frequencies of different storage elements in an array are varied using different shape anisotropies, it may be possible to address each element selectively by controlling the frequency of the RF pulse, thereby eliminating the need to include a select transistor with every storage bit.

Acknowledgements

We thank P. M. Braganca, Mark Field, and Liesl Folks for discussions. We acknowledge support from the Office of Naval Research, DARPA, and the NSF/NSEC program through the Cornell Center for Nanoscale Systems. We also acknowledge NSF support through use of the Cornell Nanofabrication Facility/NNIN and the Cornell Center for Materials Research facilities.

Author contribution

YTC performed the measurement and data analysis. JCS designed the pulse generation circuit and the measurement program. CW assisted in making the measurements. KVT helped in the device fabrication. All of the authors contributed to the analysis and the preparation of the manuscript.

Competing financial interests

The authors declare no competing financial interests.

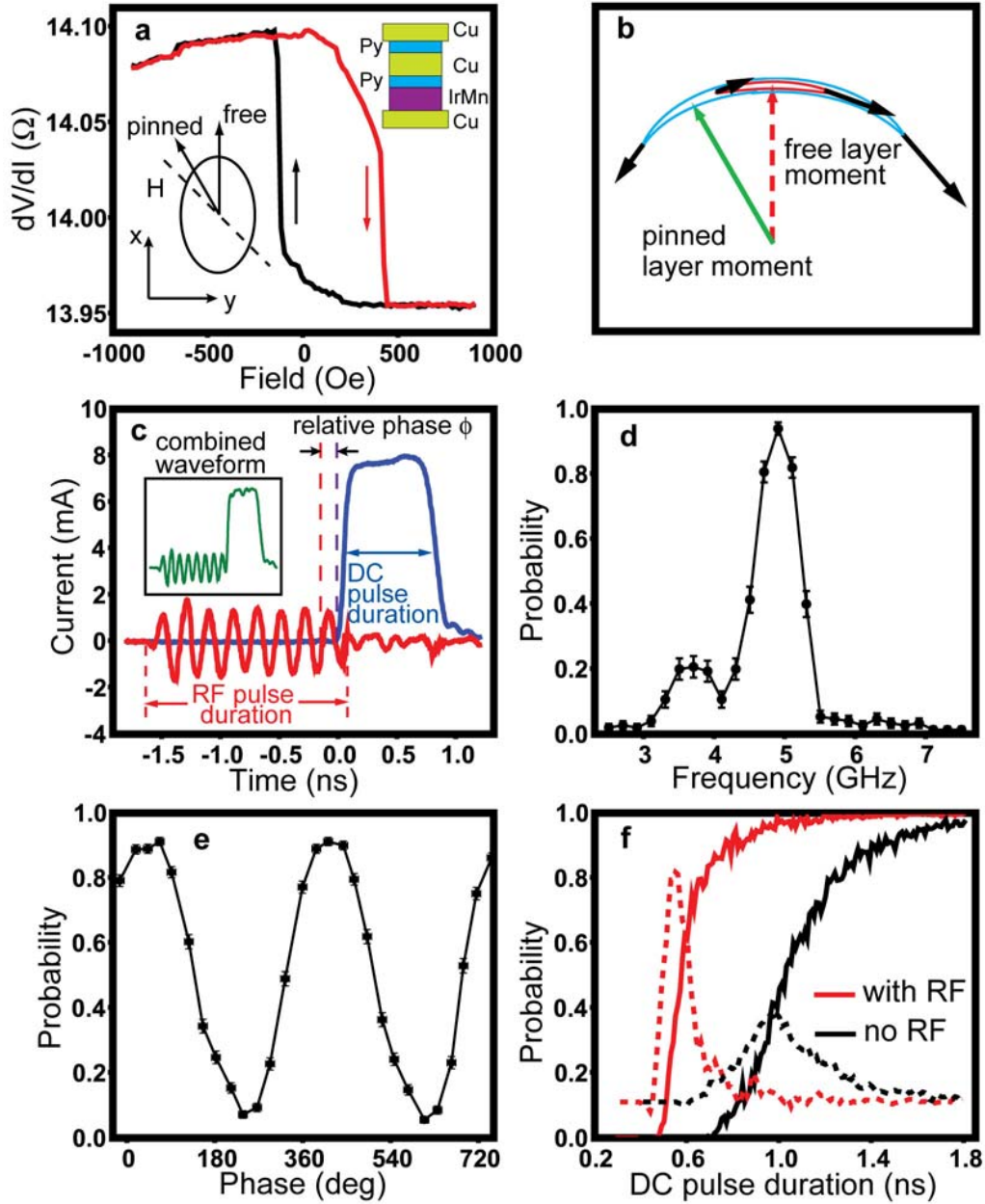

**Figure 1 Enhancement of magnetic switching using microwave current pulses at a background temperature of 20 K. a,** Resistance measured as a function of field applied along the exchange bias direction. Upper right inset: layer structure of the samples. Lower left inset: orientation of the free and pinned layer moments for the low-resistance equilibrium state. **b,** For a precession amplitude less than the offset angle between the pinned layer moment and the free layer precession axis (red trajectory), a DC current

produces a spin torque (black arrows) that increases the precession amplitude over half the cycle but decreases it for the other half. Only when the precession amplitude is greater than the offset angle (blue trajectory), will the spin torque from a DC current increase the precession amplitude on both halves of the precession cycle. **c,** Examples of our excitation waveform, which combines microwave-frequency and square-wave pulses. $\phi$ is defined as the phase of the microwave signal at the onset time of the square pulse. **d,** Switching probability as a function of frequency, after optimizing the phase $\phi$. RF pulse parameters: 1.4 mA, 1.7 ns; square pulse parameters: 7.7 mA, 0.7 ns. **e,** Switching probability as a function of the phase $\phi$ for $f = 4.9$ GHz, using the same pulse parameters as in **d**. **f,** Solid lines: switching probability as a function of square pulse duration without a microwave pulse (black curve) and with a 1.4 mA, 0.3 ns, $\phi=50°$ microwave pulse (red curve). The square-pulse amplitudes used are both 7.7 mA. Dotted lines: corresponding distributions of switching pulse duration, corresponding to the derivative of the switching probability versus pulse length.

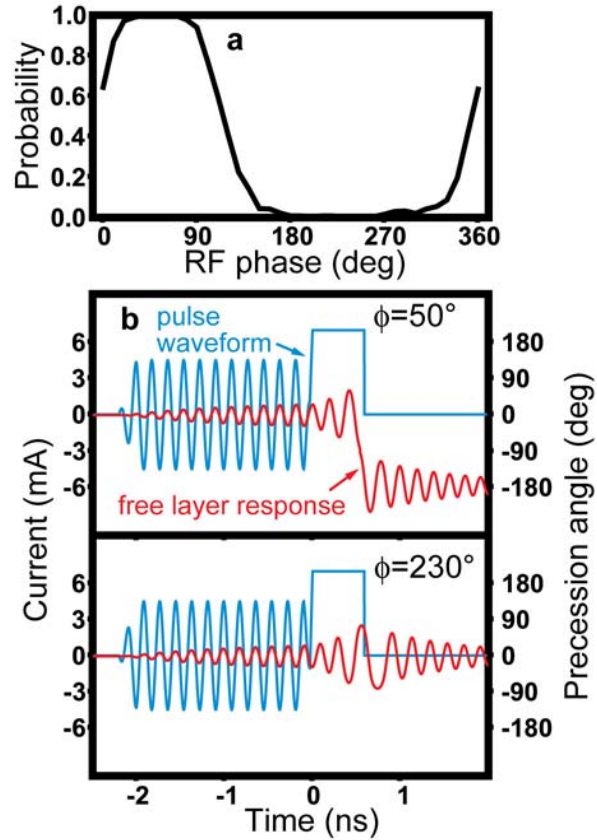

**Figure 2 Macrospin simulations of resonantly-enhanced spin-torque switching at 20 K. a,** Dependence of the switching probability on the phase $\phi$ for a 4.5 mA, 2 ns RF pulse with a 7 mA, 0.6 ns square-wave pulse. **b,** Free layer moment precession in response to the current pulse. Blue curves: Applied current pulse. Red curves: Resulting precession angle of the free-layer moment. The top panel shows for the optimum phase for switching, $\phi = 50°$. The bottom panel shows for the least-optimum phase, $\phi = 230°$. In both cases, the precession of the free layer is initially phase-locked to the applied current and the precession amplitude grows with time. For $\phi = 50°$ the square pulse is applied at the correct moment to optimally enhance the precession amplitude, leading to efficient switching, while for $\phi = 230°$ the torque from the square pulse decreases the precession amplitude over the first half-cycle of precession after the onset of the square pulse and switching is suppressed.

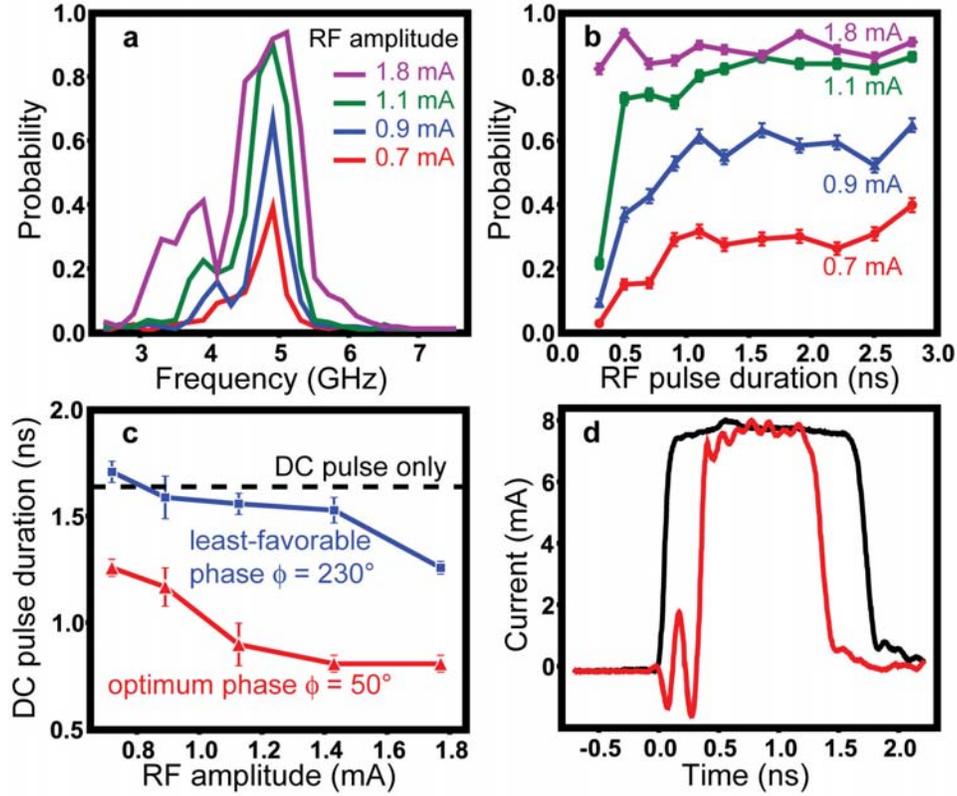

**Figure 3 Dependence of switching on the parameters of the microwave pulse at a background temperature of 20 K. a,** Switching probability versus frequency at the optimum phase $\phi$, measured using RF pulse amplitudes from 0.7 mA to 1.8 mA. The RF pulse duration is 1.7 ns and the square pulse parameters are 7.7 mA, 0.7 ns. **b,** Switching probability as a function of RF pulse duration measured at different RF pulse amplitudes from 0.7 mA to 1.8 mA. The square pulse parameters are the same as in **a**. **c,** The square pulse duration required for a 95% switching probability as a function of RF amplitude. Red curve: for the optimum phase $\phi$ for switching; blue curve: for the least-favorable phase. The black dotted horizontal line indicates the pulse duration required to give 95% switching probability for a square pulse alone. The RF pulse duration is 1.7 ns. The square pulse amplitude is 7.7 mA. **d,** Waveforms for two pulses that both give a 95% switching probability.

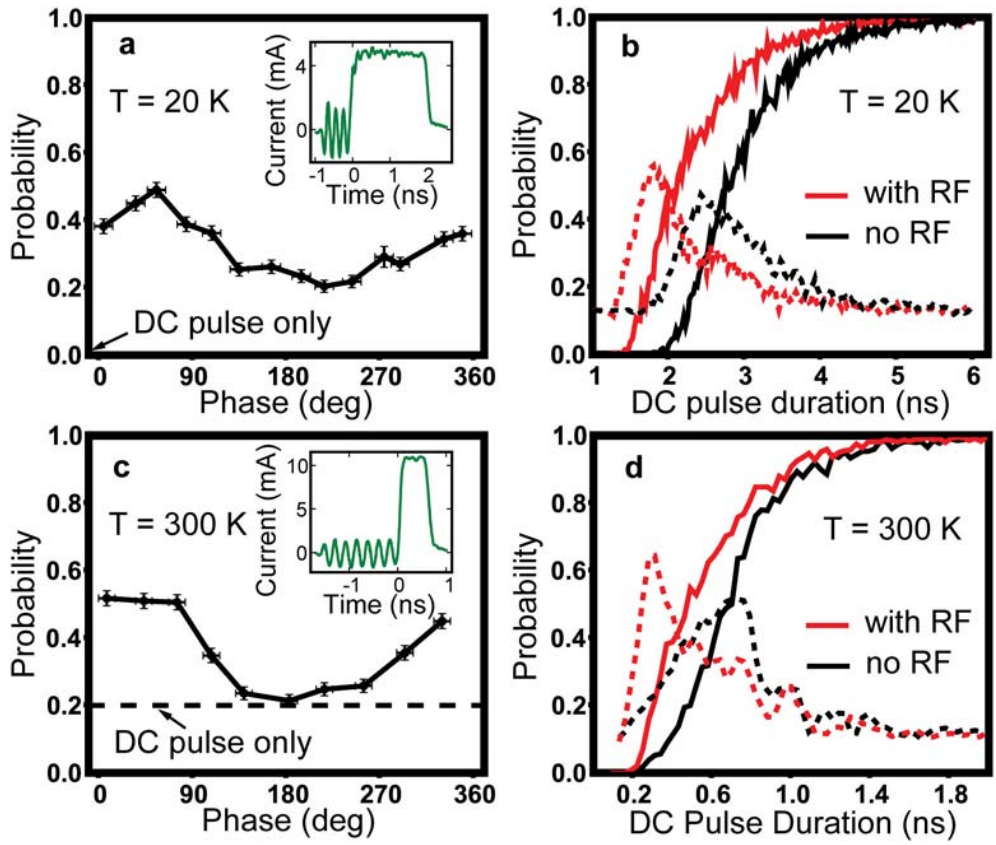

**Figure 4 Effects of thermal fluctuations on resonantly-enhanced switching. a,** Switching probability as a function of the phase $\phi$ at a background temperature of 20 K, for 1.4 mA, 0.8 ns RF pulses with 4.8 mA, 2 ns square pulses. **b,** Solid lines: Switching probability as a function of square pulse duration without an RF pulse (black) and with a 1.4 mA, 0.8 ns RF pulse (red). The square pulse amplitudes are both 4.8 mA. Dotted lines: corresponding distributions of switching pulse duration. **c,** Room-temperature switching probability as a function of the phase $\phi$, for 1.5 mA, 1.5 ns RF pulses with 10.8 mA, 0.5 ns square pulses. **d,** Solid lines: Room-temperature switching probability as a function of square pulse duration without an RF pulse (black) and with a 1.5 mA, 1.5 ns RF pulse (red). The square amplitudes are both 10.8 mA. Dotted lines: corresponding distributions of switching pulse duration.

## Supplemental Material

**Synchronized RF Pulse Circuits**

We use the mixing circuit shown in Fig. S1 to generate the waveforms for our experiment, which consist of a combination of RF and square-wave current pulses. We are able to control all of the parameters of the waveform, including the RF and square-wave pulse amplitudes, the RF phase, and the relative delay between the RF and square-wave components (within 5 ps). The RF component of the signal is generated by modulating a continuous-wave (CW) microwave source with a nanosecond-scale pulse from Pulser 1 using a mixer. A square pulse is generated by Pulser 2 and added to the RF pulse via a power divider. The combined waveform is then split using another power divider, with one signal directed to a 12.5 GHz oscilloscope for recording the waveform, and with the other directed to the sample through the RF port of a bias-tee.

In order to achieve a repeatable and controllable time delay between the RF and square pulses, and to control the phase $\phi$ of the RF signal relative to the onset time of the square pulse, the two pulsers must be precisely synchronized to the CW RF signal. We do this by using a frequency divider to generate a square-wave signal with frequency $f_{\text{trigger}} = f_{\text{RF}}/M$ (with M selected in the range 8000-12000), whose rising and falling edges are both synchronized with the CW RF signal (see Fig. S1), and then we trigger the pulsers with this signal. In order to have time to measure the sample resistance before and after each pulse, the repetition rate must be no faster than about 0.1 s. This is achieved by having the data acquisition (DAQ) computer read the square wave trigger signal and, under programmable control, emit a pulse synchronized with the square wave and lasting a time $2/f_{\text{trigger}}$ so as to gate the pulsers, enabling them to trigger. The rising edge of the gating signal is set to be synchronous with the rising edge of the signal from the frequency divider while the pulsers are set to respond to the falling edges of the triggers, thereby avoiding possible mistriggering at the rising edge of the DAQ gate signal. The pulsers do not respond to a second trigger pulse within 10 µs of the first, so to guarantee that the pulsers respond only to the first falling edge within the DAQ gate window and not the second, we make sure that the period of the signal from the frequency divider is less than 10 µs. In between excitation pulses we apply a µs-scale reset pulse consisting of a negative current with large enough amplitude to reset the sample reliably to the low-resistance magnetic state.

The phase of the RF pulse, $\phi$, relative to the onset time of the square pulse is determined by $\phi = \phi_{\text{trigger}} + 2\pi f \Delta t$, where $\phi_{\text{trigger}}$ is the phase of the RF signal at the time of the triggering edge (not tunable) and $\Delta t$ is the time delay between the triggering edge and the moment when the output square pulse arrives at the IF port of the mixer. $\Delta t$ is determined by the pulser and the length of the transmission cable, and is kept fixed during our experiment. We control $\phi$ by varying the frequency of the RF signal slightly. In our typical set-up, a frequency change of only 6.4 MHz changes $\phi$ by 360°. Our resonance linewidths are on the order of 100's of MHz, so we are able to tune $\phi$ fully using a range of frequency that is negligible compared to the linewidth.

**Measurement of switching probability**

The procedure to measure the probability of switching from the low resistance (LR) state to the high resistance (HR) state is the following. After the reset pulse is applied, we measure the differential resistance of the sample, $R_{\text{before}}$, to check the sample state and make sure that $R_{\text{before}} = R_{\text{LR}}$. Next, a gating signal from the DAQ computer is generated to enable the triggers on the pulsers, resulting in the generation of the combined RF + square-wave pulse and its transmission to the sample. After this, the differential resistance of the sample, $R_{\text{after}}$, is measured. $R_{\text{after}}$ is close to either $R_{\text{LR}}$ (not switched) or $R_{\text{HR}}$ (switched), and the result is stored. These steps are repeated $N$ times for each set of pulse parameters. The switching probability is thus calculated as $P = N_{\text{switched}} / N$ where $N_{\text{switched}}$ is the number of switched events. The uncertainty is determined as $\sigma(P) = \sqrt{P(1-P)/N}$.

**Calibration of the pulse amplitude**

The pulse waveform recorded by the sampling oscilloscope is the voltage coupled to a 50 $\Omega$ termination. We need to calibrate this voltage relative to the current coupled into the sample. This calibration is done by measuring the switching probability excited by the combination of a pulse and a steady-state current in the following procedure. First the switching probability of a 2 ns long square-wave pulse is measured for different pulse amplitudes in the absence of any steady-state current. The switching probability is 0 for small amplitude and increases to 1 as we increase the pulse amplitude. We choose a reference point $V_0$ where $P_0$ is close to 50%. Then we decrease the square pulse amplitude to $V_1$ and apply a small positive steady-state current to the sample, while measuring the switching probability corresponding to this combination of square-wave pulse $V_1$ and the steady-state current. We adjust the steady-state current, $I$, until the measured switching probability reaches the reference value, $P_0$. This means the difference between pulse amplitudes $V_0$ and $V_1$ is compensated by the applied steady state current, $I$. The voltage-to-current conversion ratio is therefore $\rho = I/(V_0 - V_1)$.

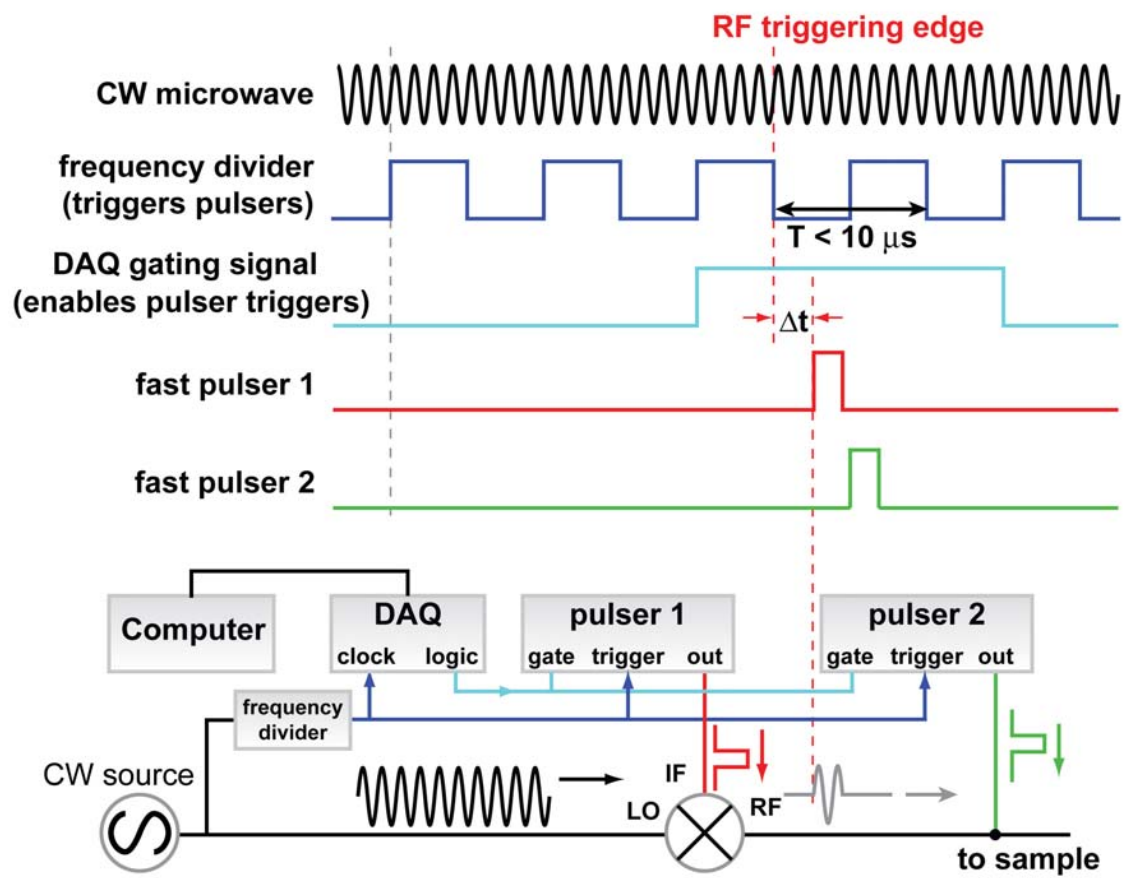

Figure S1 Circuit schematic and timing diagram for combining RF and square-wave pulses to make our excitation waveforms.

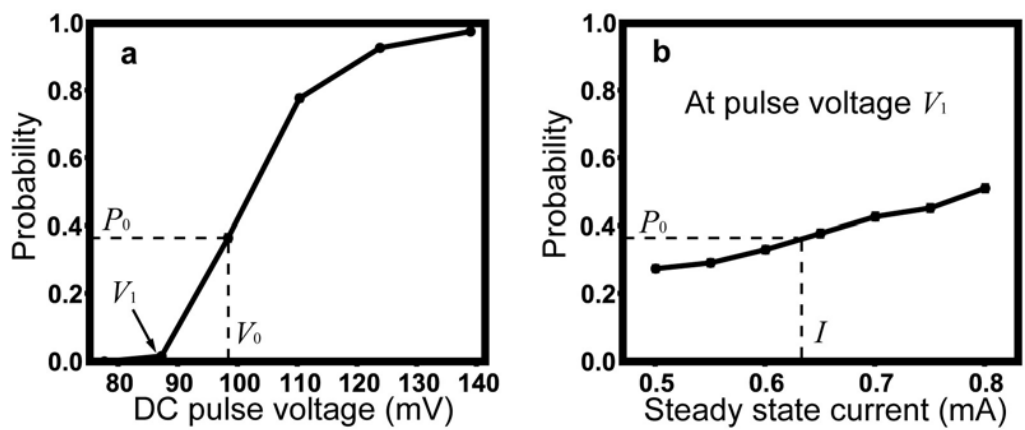

Figure S2 **a**, Switching probability as a function of pulse amplitude. $V_0$ is chosen as a reference point, with $P_0$ close to 50%. **b**, Switching probability at pulse voltage $V_1$ as a function of a steady-state background current, $I$. When $I$ is large enough to return the switching probability to $P_0$, the steady-state current is equal to the difference in the currents generated at the sample by the pulses $V_0$ and $V_1$.

26